\documentclass[epj]{webofc}
\usepackage[varg]{txfonts}   
\wocname{EPJ Web of Conferences} 
\woctitle{CONF12}
\usepackage{amssymb}
\usepackage{graphicx}
\usepackage{dcolumn}
\usepackage{bm}
\usepackage{epsfig}
\usepackage{amssymb}
\usepackage{color}
\usepackage{datetime}
\usepackage{wasysym}
\usepackage{slashed}
\usepackage[mathscr]{euscript}

\newcommand{\bal}{\begin{align}}
\newcommand{\eal}{\end{align}}
\newcommand{\beq}{\begin{eqnarray}}
\newcommand{\eeq}{\end{eqnarray}}
\newcommand{\nneeq}{\nonumber \end{eqnarray}}

\newcommand{\nn}{\nonumber \\}
\newcommand{\es}{& = &}
\newcommand{\rs}{\, = \,}

\newcommand{\cM}{ {\cal M} }
\newcommand{\cH}{ {\cal H} }

\newcommand{\cG}{ {\cal G} }
\newcommand{\cT}{ {\cal T} }

\newcommand{\cU}{ {\cal U} }
\newcommand{\cL}{ {\cal L} }

\newcommand{\cY}{ {\cal Y} }

\newcommand{\h}{ {1 \over 2} }

\begin{document}
\selectlanguage{english}
\title{Asymptotic freedom in the Hamiltonian approach to binding of color}
%
%

\author{Mar\'ia G\'omez-Rocha\inst{1}\fnsep\thanks{\email{gomezr@ectstar.eu}} 
}

\institute{European Centre for Theoretical Studies in Nuclear Physics and Related Areas (ECT*), Strada delle Tabarelle, 286  I-38123 Villazzano (TN) 
}

\abstract{%
We derive asymptotic freedom and the $SU(3)$ Yang-Mills $\beta$-function using the renormalization group procedure for effective particles. In this procedure, the concept of \textit{effective particles} of size $s$ is introduced. Effective particles in the Fock space build eigenstates of the effective Hamiltonian $H_s$, which is a matrix written in a basis that depend on the scale (or size) parameter $s$. The effective Hamiltonians $H_s$ and the (regularized) canonical Hamiltonian $H_{0}$ are related by a similarity transformation. We calculate the effective Hamiltonian by solving its renormalization-group equation perturbatively up to third order and calculate the running coupling from the three-gluon-vertex function in the effective Hamiltonian operator.
}
\maketitle
\section{Introduction}
\label{intro}
Quantum Chromodynamics (QCD) possesses asymptotic freedom~\citep{Gross:1973id,Politzer:1973fx}. The strength of interactions between quarks and gluons decreases at short distances, whereas the enlarging of separation between interacting particles leads to an increasing of the strong coupling constant, allowing quarks and gluons to form bound states and, eventually, preventing them from being isolated and forcing them to be confined. 

A precise explanation for the mechanism of confinement in QCD is still not known. However, if QCD is the theory of the strong interaction, it must predict the behavior of quarks and gluons both at short and large distances, exhibiting asymptotic freedom and yielding the whole hadron spectrum. Any comprehensive QCD approach is desired to provide explanation for the behavior of interacting particles both at low- and high-energy regimes. 

The approach presented here was formulated for the purpose of treating and solving quantum field theories within the Hamiltonian formalism. Starting from the classical Lagrangian density of the theory of interest one can derive an associated canonical Hamiltonian, expressed in terms of creation and annihilation operators acting on the Fock space. 
Such a Hamiltonian operator poses several difficulties. 
The canonical Hamiltonian is divergent and needs regularization. Furthermore, its eigenstates are superpositions of infinitely many Fock states. In this work we will consider the case of QCD without quarks. When one attempts e.g., to formulate the bound-state problem for a glueball state $|\mathbf{G}\rangle$, there is no constraint, in principle, that limits the number of Fock sectors:
\begin{eqnarray}
|\mathbf{G}\rangle \es |gg\rangle + |ggg\rangle + |gggg\rangle + \dots \ . \label{FockG}
\end{eqnarray}
This makes the bound-state problem in QCD very difficult to handle, if not impossible.

The renormalization group procedure for effective particles (RGPEP), which is going to be used here, faces this problem by introducing the concept of \textit{effective particles}. The key idea of the RGPEP is that it is possible to find an effective Hamiltonian which is related to the (regularized) canonical one by a  similarity transformation. Both Hamiltonians have the same spectrum but the effective one is written in a basis in which for a certain momentum-scale parameter $\lambda$ the number of non negligible Fock components in Eq.~(\ref{FockG}) is small, in such a way that the bound state equation can be treated and hopefully solved numerically. The notion of effective particles is also used to explain the different behavior of interacting particles at different energy scales.

The RGPEP stems form the similarity renormalization group  (SGR) procedure~\cite{Glazek:1993rc,Wilson:1994fk} and uses the Dirac's front form (FF) of Hamiltonian dynamics~\citep{Dirac}. This form has certain advantages. For example, there are no terms with only creation or only annihilation operators in the Hamiltonian and thus, the vacuum state $|0\rangle$ is an eigenstate of the the Hamiltonian with eigenvalue 0.

In the next section we present the general steps of the method of calculation, which is applied in particular to SU(3) Yang-Mills theory. In Section~\ref{SecVertex} we focus on the contribution of the effective Hamiltonian to the three-gluon vertex. In Section~\ref{SecRunningg} we show and discuss our result for the Hamiltonian running coupling. Conclusions are given in Section~\ref{Summary}.

\section{The method of calculation}
\label{sec-2}

\subsection{Canonical Hamiltonian}

We start from the classical Lagrangian density of the corresponding theory, which in this case is the SU(3) Yang-Mills Lagrangian:
\begin{eqnarray}
\cL_{YM} \es - \h \text{tr} F^{\mu\nu}F_{\mu\nu} \ ,
\end{eqnarray}
with the gluon field strength tensor being $F^{\mu\nu}= \partial^\mu A^\nu - \partial^\nu A^\mu + ig[A^\mu,A^\nu]$, $A^\mu = A^{a\mu}t^a$, $[t^a,t^b]=if^{abc}t^c$ and $\text{tr}\,  t^a t^b=\delta^{ab}/2$ .
The Noether theorem provides the associated energy-momentum tensor,
\begin{eqnarray}
\cT^{\mu \nu} \es
-F^{a \mu \alpha} \partial^\nu A^a_\alpha + g^{\mu \nu} F^{a \alpha
\beta} F^a_{\alpha \beta}/4 \ .
\end{eqnarray}
We work within the front-form of dynamics and use light-front coordinates (cf. Ref.~\citep{Brodsky-Pauli-Pinsky}), therefore $A^\mu=(A^+,A^-,A^\perp)$. In the gauge $A^+=0$, the Lagrange equations leads to the constraint
\begin{eqnarray}
A^- \es 
{ 1 \over \partial^+ } \, 2 \, \partial^\perp A^\perp 
- { 2 \over \partial^{ + \, 2} } \ 
ig \, [ \partial^+ A^\perp, A^\perp] \ .
\end{eqnarray}
With the initial conditions set on the hypersurface $x^+ = x^0 + x^3 = 0$,
the FF energy is given by
\begin{eqnarray}
H_{YM} \ = \ P^- \es {1 \over 2}\int dx^- d^2 x^\perp {\cal H}\, |_{x^+=0} \quad ,
\label{Pm}
\end{eqnarray}
with
\begin{eqnarray}
\cH \es T^{+-} \ = \ {\cal H}_{A^2} + {\cal H}_{A^3} + {\cal H}_{A^4} + {\cal
H}_{[\partial A A]^2} \ ,
\end{eqnarray}
and
\beq
\label{HA2}
{\cal H}_{A^2} \es - {1\over 2} A^{\perp a } (\partial^\perp)^2 A^{\perp a} \  , \qquad
{\cal H}_{A^3} \ = \   g \, i\partial_\alpha A_\beta^a [A^\alpha,A^\beta]^a  \ , \nn
\label{HA4}
{\cal H}_{A^4} \es  - {1\over 4} g^2 \, [A_\alpha,A_\beta]^a[A^\alpha,A^\beta]^a  \ , \quad
\label{HA2A2}
{\cal H}_{[\partial A A]^2} \ = \   {1\over 2}g^2 \,
[i\partial^+A^\perp,A^\perp]^a {1 \over (i\partial^+)^2 }
[i\partial^+A^\perp,A^\perp]^a \ .
\label{HAs}
\eeq
At the quantization surface $x^+=0$, the quantum four-vector gluon field is defined by\footnote{In the sequel we will drop the hats in particle operators, for simplicity.}
\begin{eqnarray}
\hat A^\mu \es \sum_{\sigma c} \int [k] \left[ t^c \varepsilon^\mu_{k\sigma}
\hat a_{k\sigma c} e^{-ikx} + t^c \varepsilon^{\mu *}_{k\sigma}
\hat a^\dagger_{k\sigma c} e^{ikx}\right]_{x^+=0} \ ,
\label{Amu}
\end{eqnarray}
with the shorthand notation for the integration measure $[k] = \theta(k^+)
dk^+ d^2 k^\perp/(16\pi^3 k^+) $
and the gluon polarization four-vector $\varepsilon^\mu_{k\sigma} 
= (\varepsilon^+_{k\sigma}=0, \varepsilon^-_{k\sigma} 
= 2k^\perp \varepsilon^\perp_\sigma/k^+, 
\varepsilon^\perp_\sigma)$. The indices $\sigma$ and $c$ stand for spin polarization and color, respectively. The creation and annihilation operators satisfy the commutation relation:
\begin{eqnarray}
\left[ a_{k\sigma c}, a^\dagger_{k'\sigma' c'} \right] 
\es 
k^+
\tilde \delta(k - k') \,\, \delta^{\sigma \sigma'}
\, \delta^{c c'} \ ,
\end{eqnarray}
with $\tilde \delta(p)=16\pi^3\delta(p^+)\delta(p^\perp)$. Replacing (\ref{Amu}) in (\ref{HAs}) and integrating (\ref{Pm}) one obtains the quantum canonical Hamiltonian which is written in terms of creation and annihilation operators. Such expressions are given in detail in Ref.~\citep{AF-GomezRocha-Glazek}.

\subsection{Regularization}

The Yang-Mills quantum Hamiltonian requires regularization. We adopt the same regularization procedure that was used in Ref.~\citep{Glazek2000}. Every interaction term in the Hamiltonian is labeled by its momentum quantum numbers $p^\perp$ and $p^+$; the total momentum annihilated in a vertex is labeled by $P^\perp$ and $P^+$; the relative transverse momentum is given by $\kappa^\perp = p^\perp - x P^\perp$, and the longitudinal momentum fraction is defined as $x=p^+/P^+$.
Every creation and annihilation operator of any momentum $p$ in every term is multiplied by the regulating function
\begin{eqnarray}
r_{\Delta \delta}(\kappa^\perp, x) 
\es 
\exp{(- \kappa^{\perp 2}/\Delta^2)} \, r_\delta(x) \, \theta(x) \ .
\end{eqnarray}
The first factor regulates ultraviolet divergences appearing at large $\kappa^{\perp 2}$. The Gaussian function does not allow the change of any gluon relative transverse momentum $\kappa^\perp$ to exceed the large cutoff $\Delta$. The second factor, $r_\delta(x)$, regulates divergences appearing due to small-$x$ in denominators. Its specific form  will be given later. 

The regularized canonical Hamiltonian, together with counterterms, provide the initial condition for the RGPEP equation.

\subsection{Renormalization group and effective particles}

The RGPEP uses the concept of \textit{effective particles}. The key idea is that the initial Hamiltonian $H_0$ which is written in terms of bare creation and annihilation operator of pointlike particles $a_0^\dagger$ and $a_0$ can be re-expressed in terms of creation and annihilation operators of effective particles of size $s$, $a_s^\dagger$ and $a_s$, by means of a renormalization group transformation. It is also common to use the momentum parameter $\lambda=1/s$ which distinguishes different kinds of gluons according to the rule that \textit{effective gluons of type $\lambda$ can change their relative motion kinetic energy through a single effective interaction by no more than about $\lambda$}~\citep{Glazek2000}. 
In the following, it will be convenient to use the parameter $t=s^4=1/\lambda^4$. 
Bare and effective particle operators are related by the unitary transformation\footnote{The details and the meaning of $\cH_f$ and $\cH_{Ps}$ are explained in the sequel and in Appendix~\ref{DetailsRGPEP}}
\begin{eqnarray}
a_t \es \cU_t \, a_0 \, \cU_t^\dagger \ , 
\qquad \text{with} \qquad
\cU_t
\ = \ 
T \exp{ \left( - \int_0^t d\tau \, [ \cH_f, \cH_{Ps} ] 
\right) } \ . \label{Ut}
\label{aU}
\end{eqnarray}
The RGPEP is based on the condition that the effective Hamiltonian cannot depend on the renormalization group parameter:
\begin{eqnarray}
H_t(a_t) \es H_0 (a_0) \ . \label{condition}
\end{eqnarray}
The relation (\ref{aU}) implies that $\cH_t \equiv H_t(a_0) = \cU_t^\dagger H_0 (a_0)\cU_s$. Differentiation of the latter with respect to $t$ leads to the RGPEP equation:
\begin{eqnarray}
\cH_t' 
\es \left[ [ \cH_f, \cH_{Pt} ] ,
\cH_t \right] \ .
\label{RGPEP}
\end{eqnarray}
$\cH_f(a_0)$ is the kinetic therm of the Hamiltonian and $\cH_{Pt}$ is defined in terms of $\cH_t$ as given in the Appendix. 
In order to solve the RGPEP equation for the Yang-Mills theory, we express the (unknown) effective Hamiltonian as a perturbative expansion in powers of $g$ up to third order 
\begin{eqnarray}
\cH_t
&=&
\cH_{f} + g\cH_{t1} + g^2\cH_{t2} + g^3\cH_{t3} \ .
\label{Hs}
\end{eqnarray}
Replacing this in (\ref{RGPEP}), and collecting powers of $g$ the equation can be expressed and solved order by order following the steps of  Ref.~\citep{Glazek2012}:
\begin{eqnarray}
\cH_f' \es 0 \ , \label{H0}\\
g\cH_{t\,1}' 
\es \left[\left[ \cH_f, g\cH_{1P t}\right],\cH_0\right] \ , \label{H1}\\
g^2\cH_{t\,2}' 
\es \left[\left[ \cH_f, g^2\cH_{2P t}\right], \cH_f\right] + \left[\left[ \cH_f, g\cH_{1P t}\right],g \cH_{1 t}\right] \ , \label{H2} \\
g^3 \cH_{t\,3}' 
\es \left[\left[ \cH_f, g^3\cH_{3P t}\right], \cH_f\right] + \left[\left[ \cH_f, g^2\cH_{2P t}\right], g\cH_{1 t}\right] +
\left[\left[ \cH_f, g\cH_{1P t}\right], g^2\cH_{2 t}\right]  \ . \label{H3}
\end{eqnarray}
The initial condition for solving these differential equations is given by the fact that for $s=0$ the regularized canonical Hamiltonian plus counterterm must be recovered. The latter are such that any remaining cutoff dependence must be canceled. This condition can be expressed by
\begin{eqnarray}
\cH_{t=0} \es \cH^{YM}_{\Delta\delta}  + \text{CT}_{\Delta\delta} \ ,
\end{eqnarray}
where $\text{CT}_{\Delta\delta}$ stands for counterters. The expressions for every term in (\ref{Hs}) obtained as solutions to the equations (\ref{H0})-(\ref{H3}), including counterterms, can be found in detail in Ref.~\citep{AF-GomezRocha-Glazek}. We will focus on the terms that contribute to the three-gluon vertex.

\section{The three-gluon vertex}
\label{SecVertex}

\begin{figure}[h]
\includegraphics[scale=0.7]{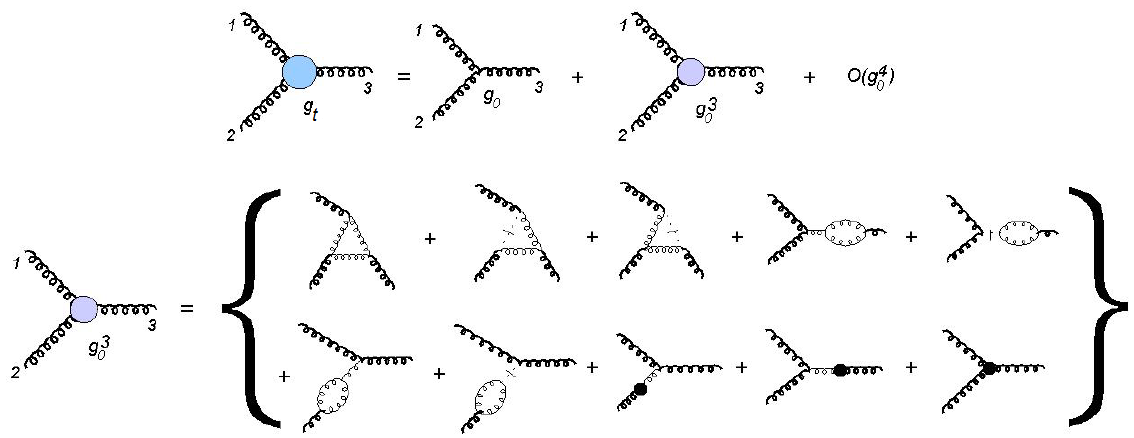}
\caption{Three-gluon vertex contributions resulting from solving the RGPEP equations perturbatively. Thick lines represent effective particle operators in Eq.~(\ref{Ys}). Thin lines represent particle operators that were present in the products of Hamiltonian terms and that are eliminated as a result of the normal ordering. Dashed lines with a crossed line indicate instantaneous  interactions. Finally, a black bubble in a line represents a self energy counterterm, while the black bubble in the vertex stands for the vertex counterterm (cf. Ref.~\cite{AF-GomezRocha-Glazek}).}\label{FigDiagrams}
\end{figure}

The three-gluon vertex Hamiltonian term contains only first- and third-order terms: 
\begin{eqnarray}
Y_t \es g \, \cH_{t1} + g^3 \, \cH_{t3} \ . \label{Ys1}
\end{eqnarray}
Third-order terms require knowledge of the solution of first- and second-order terms. 
This expression has the following form and is depicted in Fig.~\ref{FigDiagrams}.
\begin{eqnarray}
Y_t
\es 
\sum_{123}\int[123] \, \tilde \delta(k_1 + k_2 -
k_3) \,
e^{-\cM^4_{12} t} \, {\tilde r}_{\delta}(x_1)
\, \tilde Y_{t}(x_1,\kappa_{1/3}^\perp,\sigma)\, a^\dagger_{1t} a^\dagger_{2t}
a_{3t} +  H.c.
\label{Ys}
\end{eqnarray}
with ${\tilde r}_{\delta}(x)=r_{\delta}(x)r_{\delta}(1-x)$.  The labels 1, 2 and 3 stand for spin, color and momentum of each (1,2 and 3) gluon lines. The label $\sigma$ stands for spin variables and  $\cM_{12}^2={\kappa_{1/3}^{\perp 2}\over x_1x_2}$ is the invariant mass of the gluons 1 and 2.  The coefficient $\tilde Y_{t}(x_1,\kappa_{1/3}^\perp,\sigma)$ includes a three-dimensional integral over $x$ and $\kappa^\perp$ which is represented by loops in the Fig.~\ref{FigDiagrams}. This coefficient, together with the form factor $e^{-\cM^4_{12} t}$, contains the running of the vertex with the parameter $t$. 

\section{Hamiltonian running coupling and asymptotic freedom}
\label{SecRunningg}
We define the running coupling as the coefficient $\tilde Y_{t}(x_1,\kappa_{1/3}^\perp,\sigma)$ in Eq.~(\ref{Ys}) in the limit $\kappa_{1/3}^\perp \to 0$. In this limit, the vertex function has the form
\begin{eqnarray}
g_t \equiv \lim_{\kappa_{1/3}\to 0} \tilde Y_{t}(x_1,\kappa_{1/3}^\perp,\sigma)
\es
g + g^3  \cY_{t}(x_1,\kappa_{1/3}^\perp,\sigma) \ ,
\end{eqnarray}
where $\cY_{t}(x_1,\kappa_{1/3}^\perp,\sigma)$ contains counterterms, which do not depend on $t$. The difference of $g_t$ at two different values of $t$ (say $t$ and $t_0$) produces the cancellation of terms independent of $t$. We demand that at a certain $t=t_0$ or, equivalently, $\lambda=\lambda_0$, the coupling constant must have the value $g_0\equiv g_{\lambda_0}$ in agreement with experiments that correspond to $\lambda=\lambda_0$. 
The resulting running coupling calculated in this way from the three-gluon vertex in the effective Hamiltonian is 
\beq
\label{gts}
g_\lambda \es
g_0 - { g^3_0 \over 48 \pi^2 }   N_c \,  \left[ 11 + h(x_1) \right]\,\ln {\lambda \over \lambda_0} \ ,
\eeq 
where 
\beq
h (x_1) \es \chi(x_1) + \chi(1-x_1) \ , \\
\chi (x_1) 
\es
6 \int_{x_1}^1 dx \  r_{\delta Y} \left[ 2/(1-x) + 1/(x-x_1) + 1/x \right] \ - \
9 \ \tilde r_{\delta} (x_1) \int_0^1dx\,r_{\delta \mu}(x)  \, \left[ {1 \over x} + { 1 \over 1 - x}\right] \ , \label{X}  \nn 
\eeq
with
\begin{eqnarray}
r_{\delta Y}(x) \es r_\delta(x) \, r_\delta(1-x) \, r_\delta(x_1/x)
\, r_\delta[(x-x_1)/x] \, r_\delta[(x-x_1)/x_2] \, r_\delta[(1-x)/x_2] \ ,\\
r_{\delta\mu}(x) \es r_\delta^2(x)r_\delta^2(1-x) \ .
\end{eqnarray}
This result is identical to the one obtained in Ref.~\citep{Glazek2000}, where the effective Hamiltonian was calculated using another generator different from $\cG_s=[ \cH_f, \cH_{Ps} ]$ in Eq.~(\ref{Ut}). It seems that the Hamiltonian running coupling calculated within RGPEP does not depend on the generator used.  

After integration of Eq.~(\ref{X}), the result (see below) does not depend explicitly on any cutoff parameter. Every $\Delta$-dependent term was canceled by counterterms. The $\delta$-cutoff dependent terms turn out to cancel among each other. 
Nevertheless, there is still a remaining finite dependence  
on the regularization. 
We have considered the following regulating functions:
\begin{eqnarray}
\label{r1}
\text a ) \ \    r_\delta (x) \rs x/(x+\delta) \ , \quad
\label{r2}
\text b ) \  \   r_\delta (x) \rs \theta(x-\delta) \ , \quad 
\label{r3}
\text c ) \ \    r_\delta (x) \rs x^\delta \, \theta(x-\epsilon) \ ,
\label{regs}
\end{eqnarray}
and the obtained $h(x_1)$ in Eq.~(\ref{gts}) for each case is:
\beq
\text a ) & &   h(x_1) \, = \, 12\left[ 3 + {1-x_1 - x_1^2 \over (1-x_1)(1-2x_1)} \ln{x_1} +
                               {(1-x_1)^2 -x_1 \over x_1(1-2x_1) } \ln{(1-x_1)} \right] \ , \label{ha}\\
\text b ) & &   h(x_1) \, = \, 12 \ln{ \, \text{min}(x_1,1-x_1)} \ , \label{hb}\\
\text c ) & &   h(x_1) \, = \, 0 \ \label{hc} . 
\eeq

\begin{figure}[h]
\centering
\sidecaption
\includegraphics[scale=0.6]{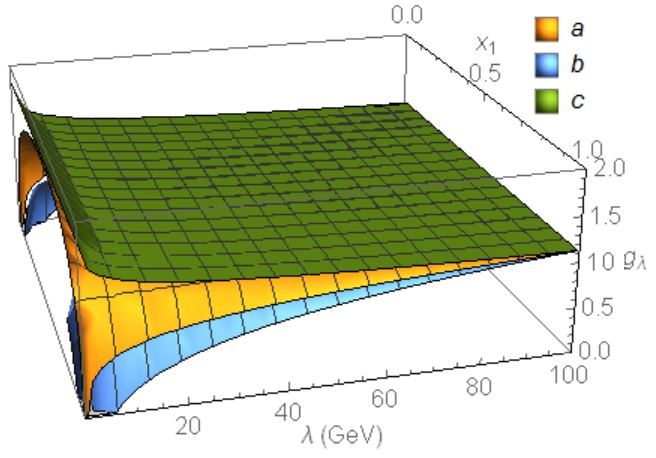}
\caption{Hamiltonian running coupling obtained using three different regularizations (cf.~Eqs.~(\ref{ha})-(\ref{hc})) as a function of $x_1$ and $\lambda$,   
for $\lambda_0=100$ GeV and $g_0=1.1$.
} 
\label{FigRunningg}
\end{figure}

\begin{figure}[h]
\centering
\sidecaption
\includegraphics[scale=0.51]{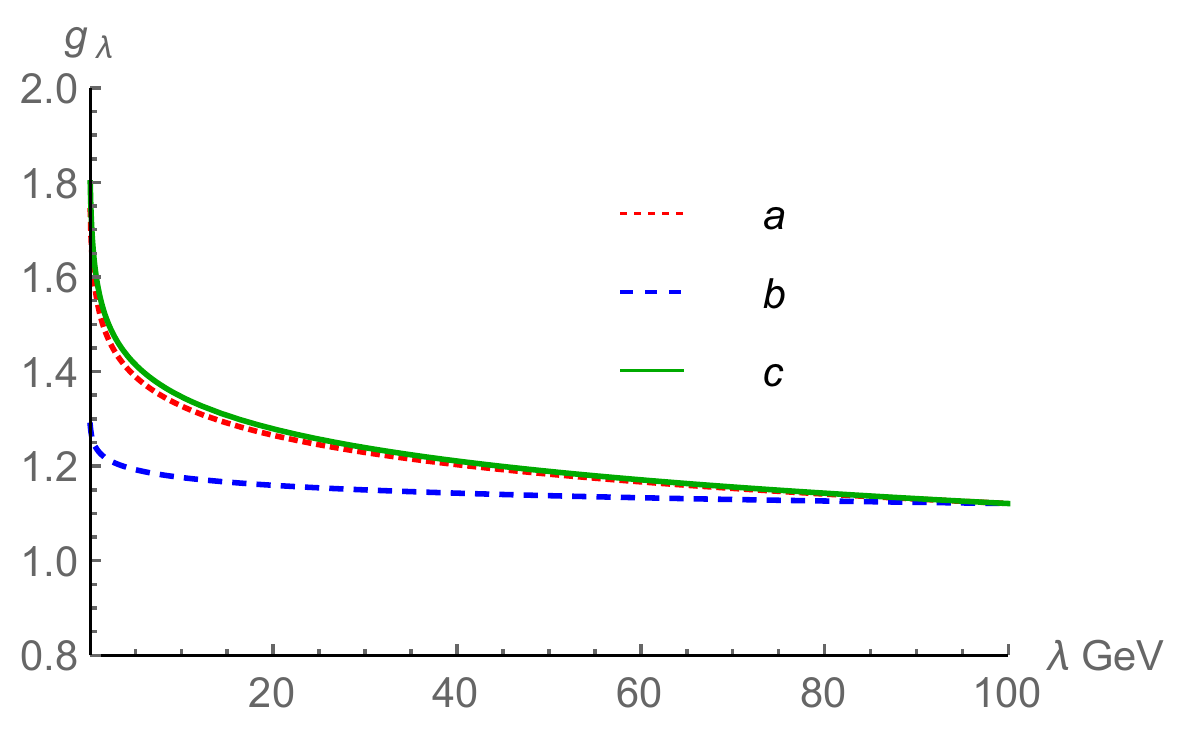}
\includegraphics[scale=0.5]{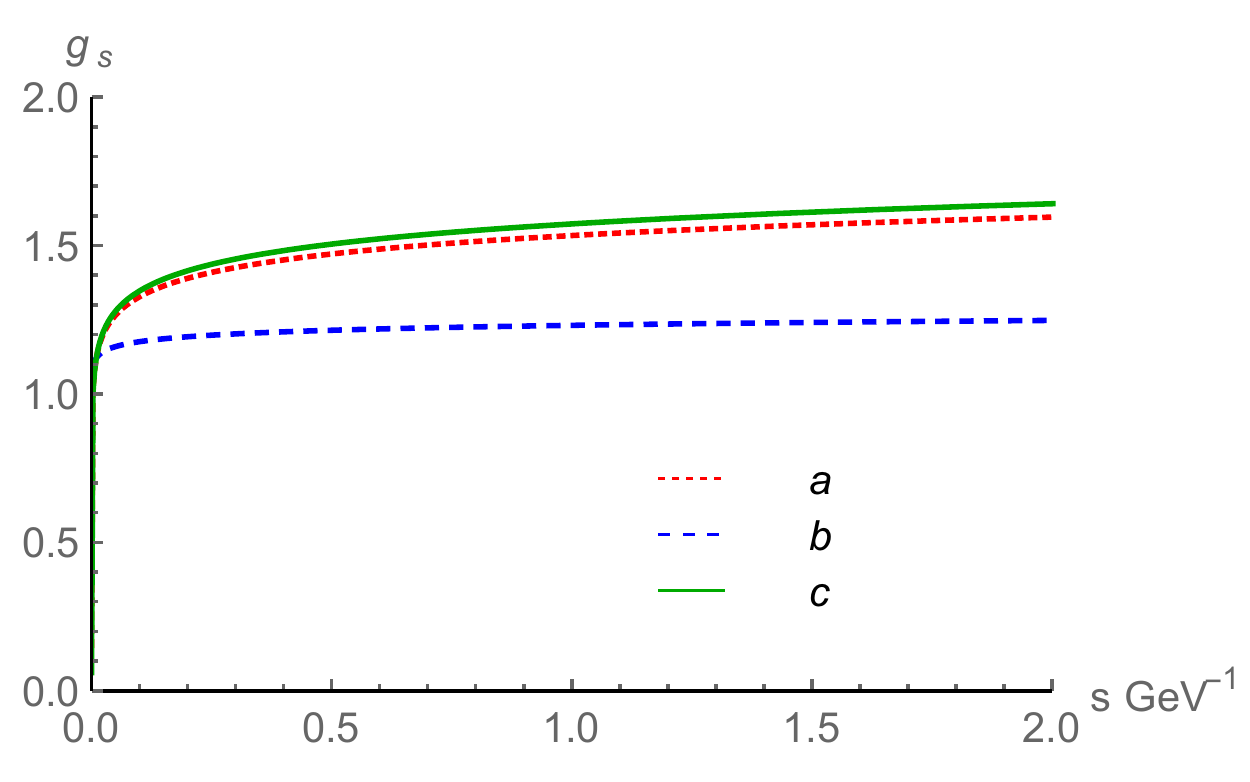}
\caption{ 
The Hamiltonian running coupling for different regularizations and $x_1=0.5$, as a function of momentum scale $\lambda$ (left) and effective size $s=1/\lambda$ (right). In both cases  
 $\lambda_0=100$ GeV and $g_0=1.1$.} 
\label{FigRunningg2D}
\end{figure}

We have plotted the running coupling as a function of $\lambda$ and $x_1$ in Fig.~\ref{FigRunningg}. One sees, especially for certain $x_1$-regions, that it is necessary to carefully account for effects due to small-$x$ regularization.  
Fig.~\ref{FigRunningg2D} shows the running coupling for $x_1=0.5$ as a function of $\lambda$ (left panel) and as a function of the size $s$  of the effective gluons (right panel). Particles that interact very weakly due to asymptotic freedom are particles of very small size, whereas larger effective particles interact strongly. This fact has been argued and sketched in a picture in Ref.~\cite{Condensates}. 

It is remarkable that for the regularization $(c)$ there is no dependence on $x_1$, and the result maches the asymptotic freedom result in Refs.~\cite{Gross:1973id,Politzer:1973fx}
if one identifies $\lambda$ with the momentum 
scale of external gluon lines in Feynman diagrams:
\beq
\label{gl}
g_\lambda \es
g_0 - { g_0^3 \over 48 \pi^2 }   N_c \,   11 \,\ln
{ \lambda \over \lambda_0} \ ,
\eeq 
and
\beq
\lambda {d \over d\lambda} \, g_\lambda
\es \beta_0 g_\lambda^3 \ , \qquad \text{with} \qquad \beta_0 \rs - { 11 N_c \over 48\pi^2 } \ .
\eeq

The dependence on the regularizations found in ($a$) and ($b$) clearly reflects the fact that the effective third-order Hamiltonian depends on the regularization. This does not necessarily mean that the full effective Hamiltonian must depend on the regularization. It is possible that cancellations occur at higher orders. Only higher order analysis of the running coupling may improve the understanding of asymptotic freedom in our Hamiltonian approach to Yang-Mills theory.

Nonetheless, it is outstanding that for regularization $(c)$ we obtain the standard result. This outcome suggests a possible direction in which one could seek a path towards the equivalence of the Euclidean Green's function calculus and Minkowskian Hamiltonian quantum field operator.

\section{Summary and conclusions}
\label{Summary}

We have derived an effective Hamiltonian for SU(3) Yang-Mills theory starting from its Lagrangian density and using the RGPEP. 
We have introduced the concept of effective gluons of size $s$ or associated momentum width $\lambda$. Finally, we have calculated the corresponding three-gluon vertex function from which we have extracted the Hamiltonian running coupling in the limit $\kappa_{1/3}^\perp\to 0$, and certain values of $x_1$.

The derived running coupling has the same asymptotic-freedom behavior as in the calculus based on renormalized Feynman diagrams. This is important from the point of view of the desired but unknown precise connection between Feynman diagrams for virtual transition amplitudes and the Hamiltonan formalism in the Minkowski spacetime. The precise behavior is not observed equally for all small-$x$ regulating functions using third-order Hamiltonians. 

In our third-order calculation we have observed a certain dependence on the regularization of small-$x$ divergences. This dependence is identical to the one observed in analogous calculations using a different RGPEP generator. 
This suggests another universal aspect of the RGPEP; namely, that the way how the running coupling depends on the size of the effective gluons is independent of the generator used. 

The studies presented here require Hamiltonian matrix elements of terms with one creation and two annihilation operators and viceversa. In order to set any physical bound state problem, such as the eigenvalue problem for a glueball, the construction of an effective Hamiltonian requires terms with the same number (at least two) of creation and annihilation operators. Then, at least fourth order RGPEP Hamiltonians are required if one wants to include the running of the coupling.

\begin{appendix}
\section{Details of the RGPEP}
\label{DetailsRGPEP}
\end{appendix}
The details of elements present in Eq.~(\ref{RGPEP}) are:
\begin{eqnarray}
\cH_f \es
\sum_i \, p_i^- \, a^\dagger_{0i} a_{0i} \ , \qquad \text{with} \qquad  p^-_i = { p_i^{\perp \, 2} \over p_i^+} \ , \\
\cH_t(a_0) \es
\sum_{n=2}^\infty \, 
\sum_{i_1, i_2, ..., i_n} \, c_t(i_1,...,i_n) \, \, a^\dagger_{0i_1}
\cdot \cdot \cdot a_{0i_n} \ ,\\
\cH_{Pt}(a_0) \es
\sum_{n=2}^\infty \, 
\sum_{i_1, i_2, ..., i_n} \, c_t(i_1,...,i_n) \, 
\left( {1 \over
2}\sum_{k=1}^n p_{i_k}^+ \right)^2 \, \, a^\dagger_{0i_1}
\cdot \cdot \cdot a_{0i_n} \ .
\end{eqnarray}
The operator $\cH_f$ is called 
the free Hamiltonian, for being the part of $\cH_0(a_0)$ that does not depend on the coupling constant. The index $i$ denotes the quantum numbers of gluons 
and $p_i^-$ is the free FF energy for the gluon 
kinematical momentum components $p_i^+$ and 
$p_i^\perp$.

The operator $\cH_{Pt}$ differs from  
$\cH_t$ in the multiplication of each and every term by the square of the total + momentum involved in a term.  

\vspace{0.5cm}
\noindent
\textbf{Acknowledgments:} Part of the development of this work was supported by the Austrian Science Fund (FWF), Project No. P25121-N27. Figure~\ref{FigDiagrams} has been made using JaxoDraw~\citep{Binosi}

\end{document}